\documentclass[preprint,12pt]{elsarticle}
\usepackage{graphicx}
\usepackage{dcolumn}
\usepackage{amssymb}

\newcommand{\beq}{\begin{equation}}
\newcommand{\eeq}{\end{equation}}
\newcommand{\beqa}{\begin{eqnarray}}
\newcommand{\eeqa}{\end{eqnarray}}

\journal
{Physics Letters B}

\begin{document}

\begin{frontmatter}
\title{Statistical interpretation of the spatial distribution of current
130 GeV gamma-ray line signal within the dark matter annihilation scenario}

\author{Rui-Zhi Yang$^{(a)}$$^{(b)}$, Qiang Yuan$^{(c)}$$^{(a)}$, Lei Feng$^{(a)}$, Yi-Zhong Fan$^{(a)}$\footnote{The corresponding author (email: yzfan@pmo.ac.cn)} and Jin Chang$^{(a)}$}

\address[a] {Key Laboratory of Dark Matter and Space Astronomy, Purple Mountain Observatory, Chinese Academy of Sciences, Nanjing
210008, China}
\address[b]{Max-Planck-Institut f{\"u}r Kernphysik, P.O. Box 103980, 69029 Heidelberg, Germany}
\address[c]{Key Laboratory of Particle Astrophysics, Institute of High Energy
Physics, Chinese Academy of Science, Beijing 100049, China
\\}
\begin{abstract}
Recently, several groups identified a tentative $\gamma$-ray line signal
with energy $\sim 130$ GeV in the central Galaxy from the Fermi-LAT data.
 Such a $\gamma$-ray line can be interpreted as the signal of dark matter annihilation.
However, the offset $\sim 220$ pc ($1.5^{\circ}$) of the
center of the most prominent signal region from the Galactic center Sgr A$^{\star}$ has been thought
to challenge the dark matter annihilation interpretation.
Considering the fact that such a 130 GeV $\gamma$-ray line signal consists
of only $\sim14$ photons, we suggest that the ``imperfect''
consistency of these photons with the expected dark matter distribution
is due to the limited statistics. The offset will be smaller as more
signal photons have been collected in the near future. Our Monte Carlo
simulation supports the above speculation.
\end{abstract}
\end{frontmatter}

High energy $\gamma$-ray line is of extreme interest in search for the
signal of dark matter (DM) annihilation or decay. Recently, via analyzing
the publicly available Fermi-LAT $\gamma$-ray data, Bringmann {\it et al.}
\cite{bringman} and Weniger \cite{weniger} found weak evidence for a
monochromatic $\gamma$-ray line with energy $\sim130$ GeV.
Later independent analyses carried out by a few groups confirmed the
existence of the 130 GeV $\gamma$-ray excess, and the signal has been
found to be even more prominent \cite{tempel,boyarsky,sumeng}. This
result can be interpreted by $\sim 130$ GeV DM annihilation, with
annihilation cross section $\langle\sigma v\rangle_{\rm \chi\chi\rightarrow
\gamma\gamma}\sim 10^{-27} {\rm cm^3~s^{-1}}$, and a cuspy density profile
such as Navarro-Frenk-White (NFW, \cite{nfw}) and Einasto \cite{einasto}.

Much attention was paid on this line signal in the community. Many models
were proposed to explain this line structure, either by DM \cite{model}
or astrophysical sources \cite{profumo,aharonian}. It was also suggested
to constrain the DM scenarios with the continuum $\gamma$-rays or
antiprotons \cite{constrain}, or to test the line postulation with
high energy resolution detectors \cite{test}. Moreover, the spectra
of the sum of cosmic ray electrons and positrons detected by ATIC and
PAMELA both showed small wiggle-like structures at $\sim 100$ GeV
\cite{atic,pamela-russia}, which could be the result of the annihilation
of $\sim 140$ GeV DM particles into electrons/positrons, in accordance
with the 130 GeV $\gamma$-ray line \cite{Feng2012}.

Among current relevant data analysis works, the morphology of the
potential line signal is still in debate \cite{tempel,boyarsky,sumeng}.
It is very attractive that Su \& Finkbeiner identified that the signal
region lies basically in the Galactic center region, and the detection
significance is quite high (exceeds $5\sigma$) \cite{sumeng}. The former
character is just expected in the DM scenario, while a confidence level
$>5\sigma$, if confirmed, is encouraging to approach a discovery/detection.
The problem is however that the signal region has a center deviating
from the Galactic centre (Sgr A$^{\star}$) considerably by a distance
$\sim 220$ pc (or angle $\sim 1.5^{\circ}$)\footnote{For the signal
region with highest significance ($\sim 4.5 \sigma$, i.e., their
central region) identified in \cite{tempel}, an offset $\sim 1.2^\circ$
was reported, too.}. The result seems to be at odds with the DM models
in which the signal region is expected to be centered at $(\ell,~b)=
(0^{\circ},~0^{\circ})$, where $(\ell,~b)$ are the Galactic longitude
and latitude, respectively. Indeed, such a puzzle has been thought to
be one of the strongest arguments against the DM origin of the $\gamma$-ray
line signal \cite{aharonian,sumeng12b}. In this Letter, considering
the fact that the current 130 GeV $\gamma$-ray line signal consists
of only $\sim14$ photons, we try to provide a statistical interpretation
of the spatial distribution.

For such a purpose we carry out the Monte Carlo simulation of the arrival
direction of the photons produced by the annihilation of DM particles.
Following \cite{sumeng} we adopt the Einasto DM density profile in this
work \cite{einasto}
\begin{equation}
\rho(r)=\rho_{\rm s}\exp\left(-\frac{2}{\alpha}\left[\left(\frac{r}
{r_{\rm s}}\right)^{\alpha}-1\right]\right),
\end{equation}
where $\alpha=0.17$, $r_{\rm s}\approx 20$ kpc and $\rho_{s}\approx 0.06$
GeV cm$^{-3}$.

The possibility of detecting one photon at the location $(\ell,~b)$ is
proportional to the $J$-factor
\begin{equation}
J\propto\int {\rm d}s \rho^2(r(s)),
\end{equation}
where $r=(s^{2}+r_\odot^{2}-2sr_\odot \cos \ell \cos b)^{1/2}$ is the
Galactocentric distance, $r_\odot \simeq 8.5$ kpc is the distance from
the Sun to the Galactic center and $s$ is the line of sight distance.

\begin{figure}
\centering
\includegraphics[width=0.84\columnwidth,angle=0]{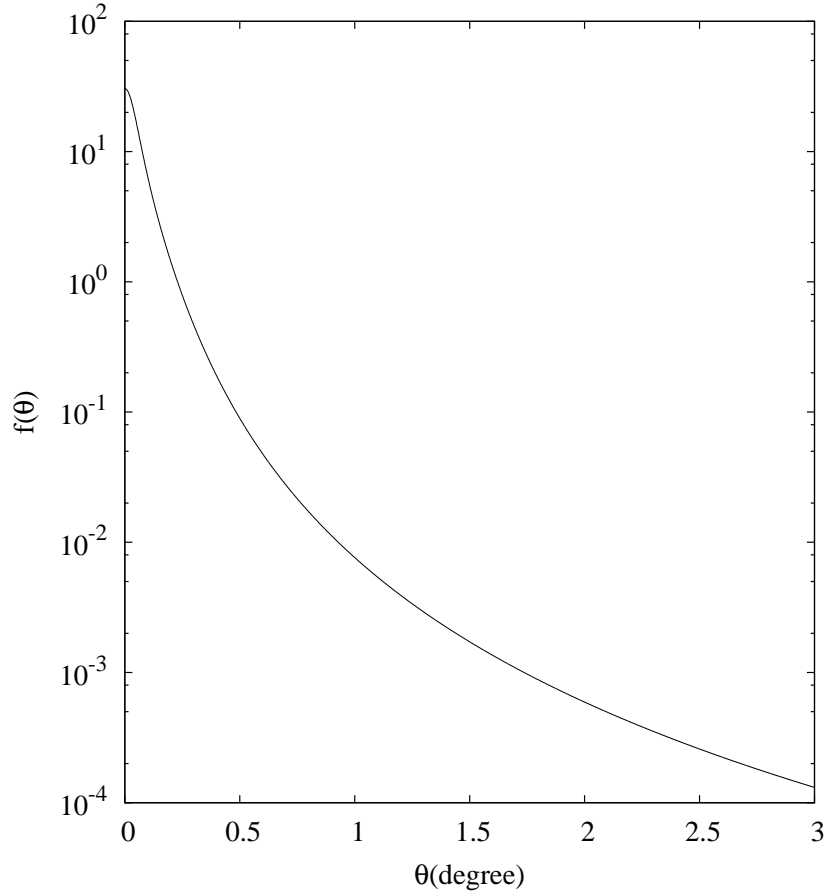}

\caption{ The inclination-angle averaged point spread function of Fermi-LAT at $130~\rm GeV$.}  \label{fig:pst}
\end{figure}

In reality the ``observed'' (reconstructed) direction of the photons will
deviate from the ``real'' location, and the deviation is a function of an
incident photon's energy and inclination angle. Such an effect is known
as the point-spread function (PSF) of the instrument. For Fermi-LAT, the
PSF function can be found on the website\footnote{http://fermi.gsfc.nasa.gov/ssc/data/analysis/documentation/Cicerone/Cicerone\_LAT\_IRFs/IRF\_PSF.html}.
In this work, we fix the photon energy to be $130~\rm GeV$ and average the
PSF in different inclination angles. The derived PSF function $f(\theta)$
is shown in Fig. \ref{fig:pst}, where the normalization $2\pi
\int{f(\theta)}d\theta=1$ has been adopted. For each photon in the
simulation, we re-generate the ``observed'' direction which may deviate
$\theta$ from the ``real'' direction with probability $f(\theta)$.

\begin{figure}
\centering
\includegraphics[width=0.42\columnwidth,angle=0]{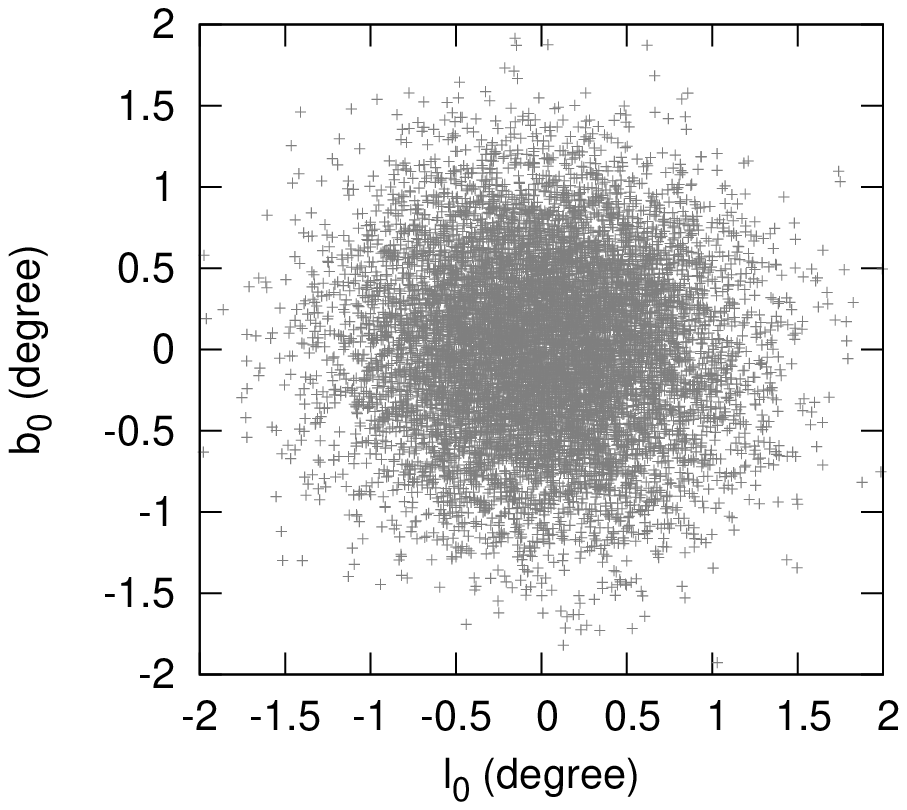}
\includegraphics[width=0.42\columnwidth,angle=0]{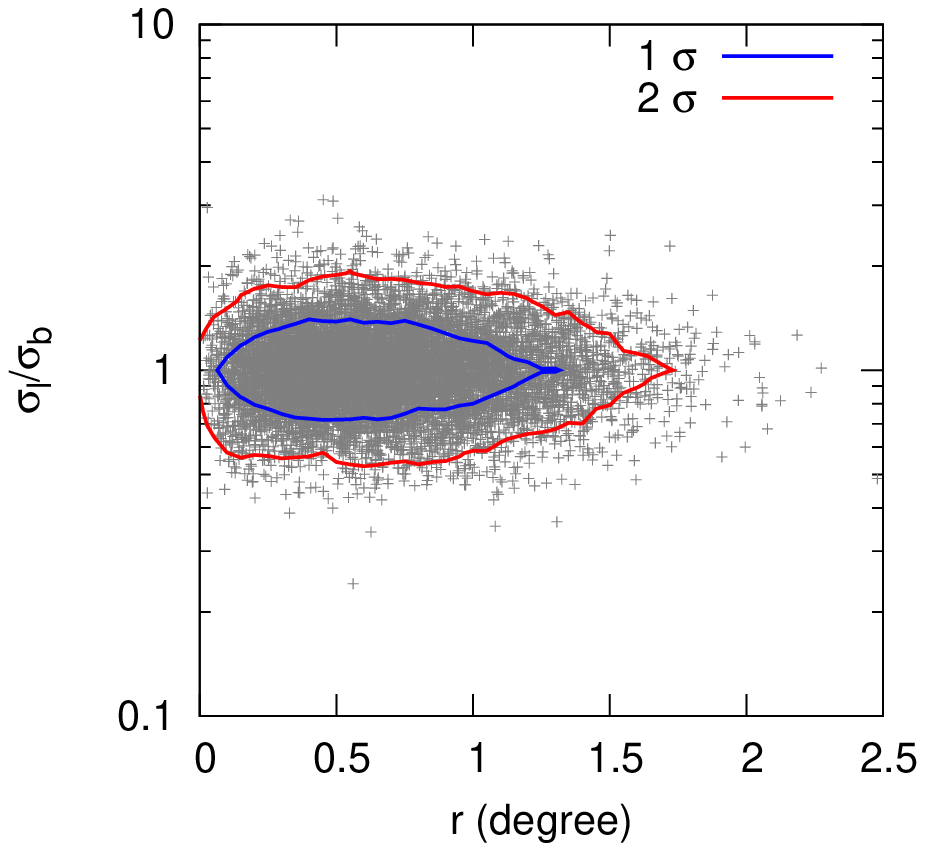}
\caption{Left: scattering plot of $(\ell_0,~b_0)$ in the $10^4$
simulations; right: scattering plot of offset angle $r$ versus the
elongation rate $\sigma_{\ell}/\sigma_b$.} \label{fig:f1}
\end{figure}

We simulate 10000 observations with $N=14$ photons each. These
photons are assumed to come from an angle $\xi\leq 5^\circ$ around the
Galactic center since all the 14 signal photons identified in
\cite{sumeng} were from such a very compact region, where
$\cos \xi=\cos \ell \cos b$ (for the Einasto DM density profile, just
about $1/4$ of the signal photons will be from $\xi\leq 5^\circ$,
implying that there are many more signal photons from larger angles,
which however may have been hidden behind the dense background).
The average center of the photons is estimated to be
\begin{equation}
\ell_0=\sum\limits_{i=1}^{N} \ell_{\rm i}/N,~~~~
b_0=\sum\limits_{i=1}^{N} b_{\rm i}/N.
\end{equation}
The distribution of the resulting $(\ell_0,~b_0)$ is presented in the
left panel of Fig. \ref{fig:f1}.

The offset of the morphology center compared with the Galactic center
is simply $r=\sqrt{\ell_0^{2}+b_0^{2}}$. We also investigate the asymmetric
property of the photon map, through defining the elongation rate
\begin{equation}
\sigma_{\ell}/\sigma_b=\sqrt{\frac{1}{N}\sum\limits_{i=1}^{N}(\ell_{\rm i}-
\ell_0)^{2}} \left/ \sqrt{\frac{1}{N}\sum\limits_{i=1}^{N}(b_{\rm i}-b_0)^{2}}
\right..
\end{equation}
In the right panel of Fig. \ref{fig:f1} we show the distribution of
$r$ and $\sigma_{\ell}/\sigma_b$. Lines in this figure present the $1\sigma$
and $2\sigma$ contours. It is shown that an offset of about $1.5^{\circ}$
revealed by the data is consistent with the canonical DM distribution
within $2\sigma$ confidence level. Specifically the probability of
$r>1.5^{\circ}$ is about $2\%$ (or $2.3\sigma$), for the case $N=14$.
Our prediction will be directly tested by the ongoing and upcoming high
energy observations. The results also show that a potential asymmetry
between the longitude and latitude directions \cite{sumeng} may also
appear due to the limited statistics.

\begin{figure}
\centering
\includegraphics[width=0.42\columnwidth,angle=0]{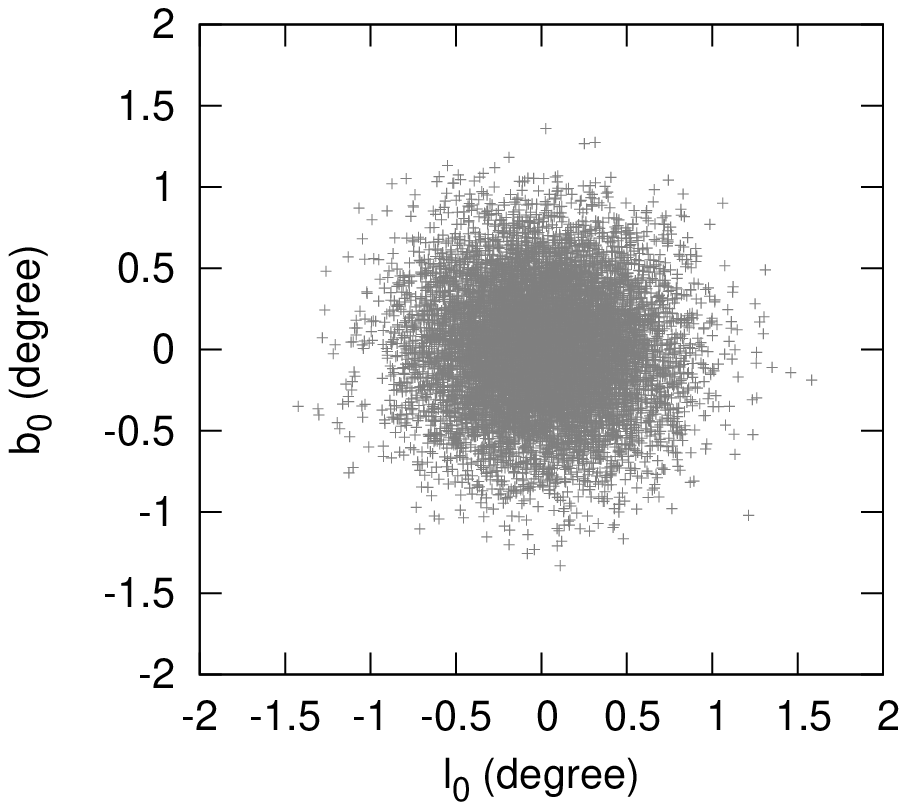}
\includegraphics[width=0.42\columnwidth,angle=0]{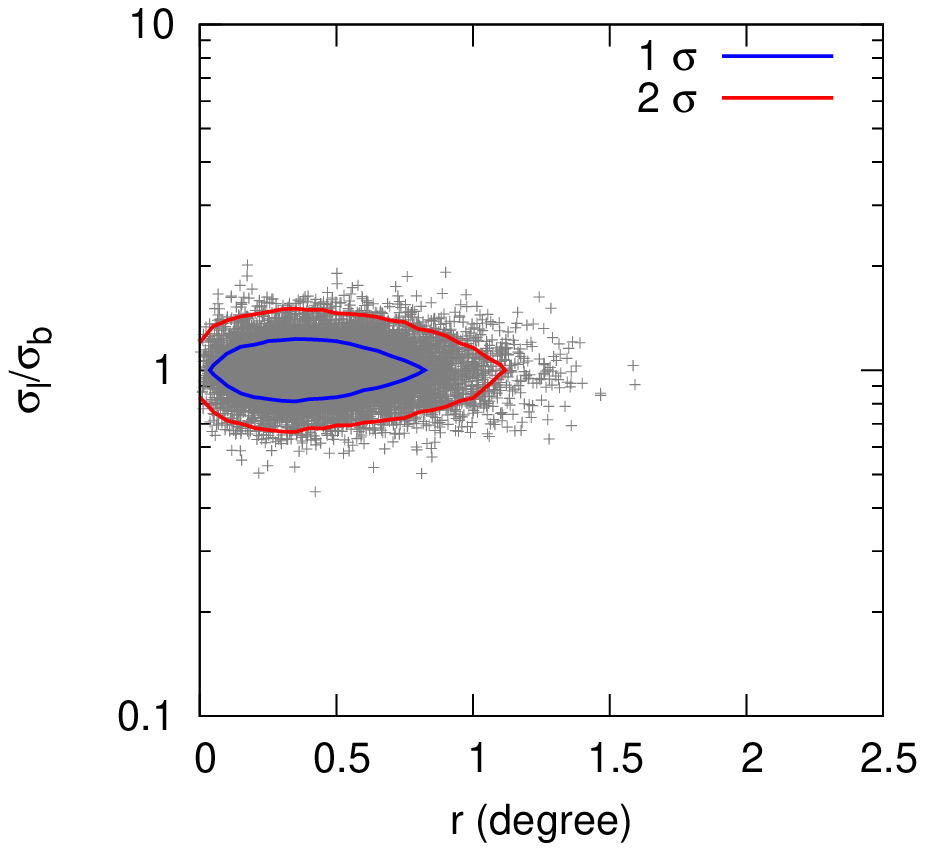}
\includegraphics[width=0.42\columnwidth,angle=0]{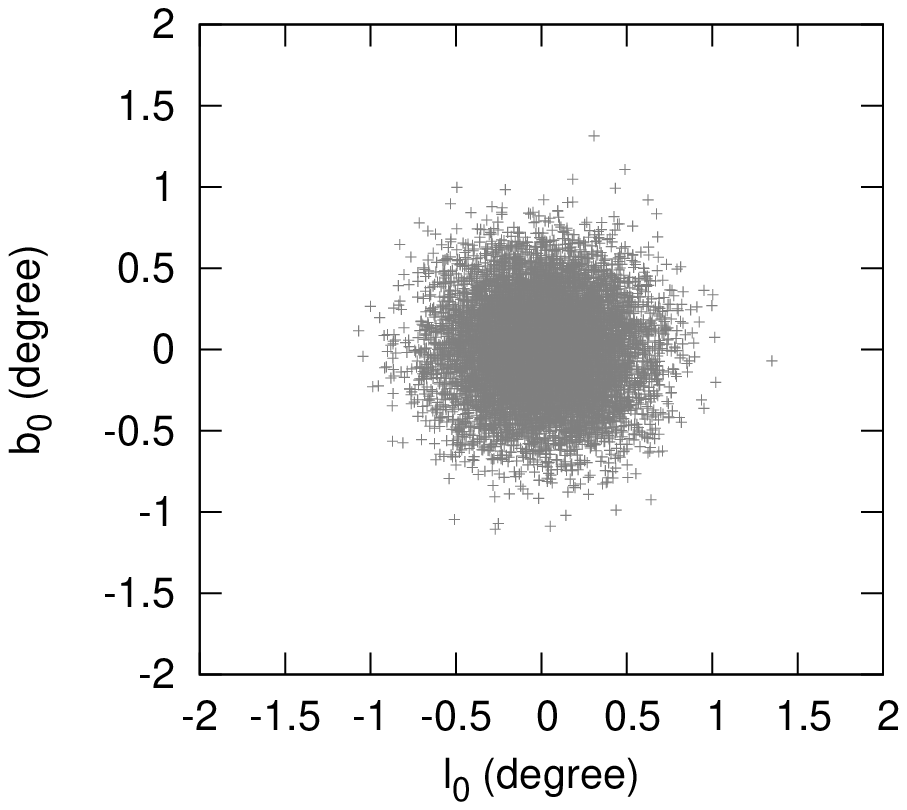}
\includegraphics[width=0.42\columnwidth,angle=0]{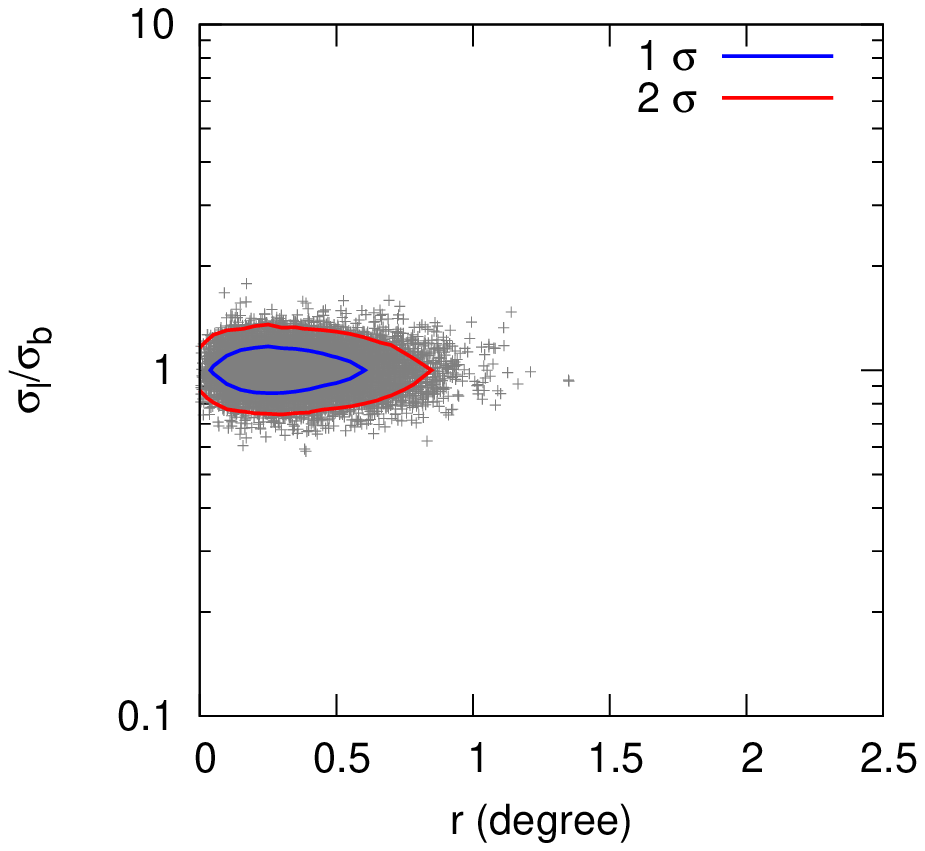}
\includegraphics[width=0.42\columnwidth,angle=0]{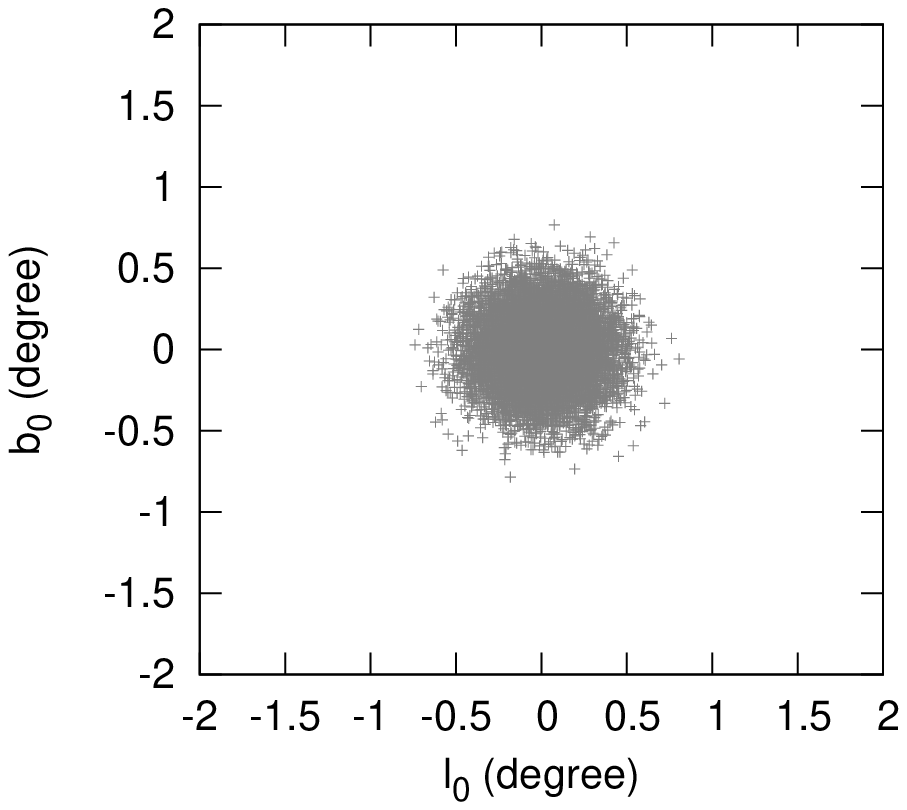}
\includegraphics[width=0.42\columnwidth,angle=0]{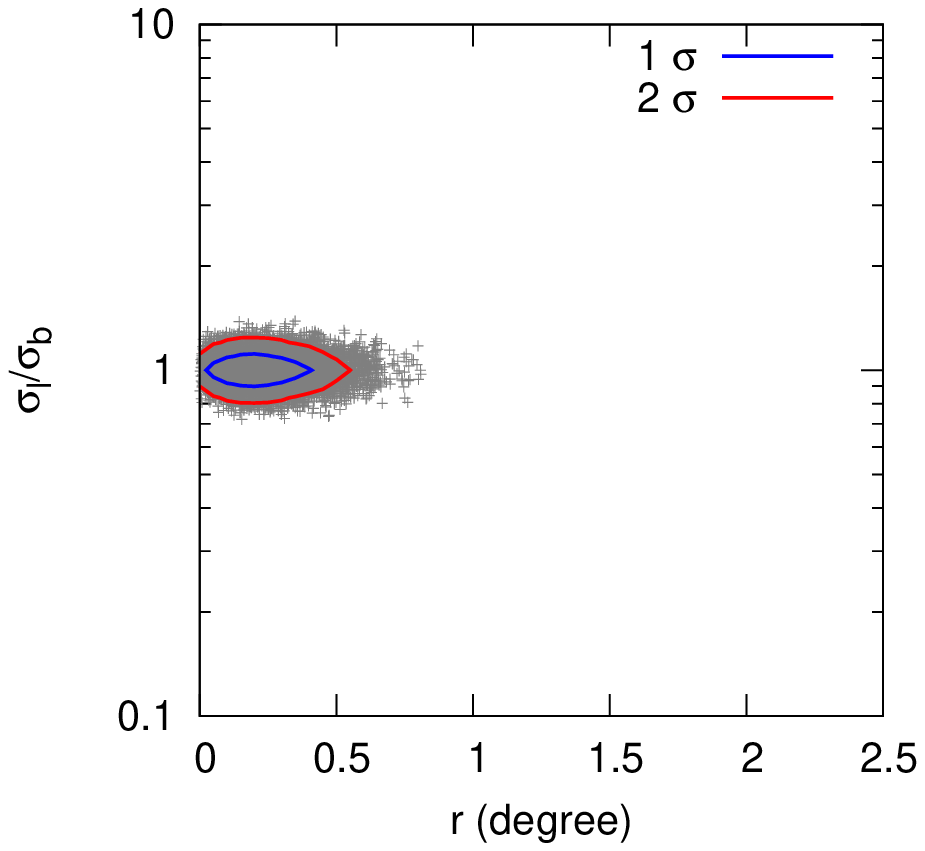}
\caption{Same as Fig. \ref{fig:f1}, but for $N=30$ (top), $N=50$ (middle)
and $N=100$ (bottom) respectively.} \label{fig:f2}
\end{figure}

With the increase of photon statistics, we would expect the deviation of
the morphology center from the real center to decrease. The resulting
distributions of the parameters for $N=30$, $50$ and $100$, still for
10000 simulations, are shown in Fig. \ref{fig:f2}. Given more and more
photons we find that the fluctuation of the morphology
center becomes smaller and smaller, as expected. The probability
for a large offset of the morphology center is accordingly much smaller.
For $N=30$ we have $P(r>1.5^{\circ})=0.02\%$. This probability decreases
to $8\times10^{-7}$ for $N=50$, and become much smaller for $N=100$.
Obviously, the morphology will be more symmetric given more photons
are detected.

In summary we have shown that the spatial distribution of 130 GeV
$\gamma$-ray line signal identified in \cite{sumeng} can be consistent
with the DM annihilation model and the offset of the signal region from
the Galactic centre is likely caused by the limited statistics.
The upcoming high energy resolution detectors such as DArk Matter Particle
Explorer (DAMPE) and CALorimetric Electron Telescope
(CALET)\footnote{http://calet.phys.lsu.edu/} will be more powerful in
identifying the line-like $\gamma$-ray signal \cite[e.g.,][]{test}.
These two detectors however have an effective area smaller than Fermi-LAT,
and the total photons detectable should also be fewer. We thus do not
expect to get a perfect coincidence of the signal region with the expected
DM distribution even in future observations. In this Letter the distribution of dark matter particles in the central Galaxy is assumed to be centered at $(\ell,~b)=(0^{\circ},~0^{\circ})$. If it is not the case an offset lager than that predicted in our simulation is likely and the current spatial distribution of the signal photons might be better reproduced.\\

{\it Acknowledgments:} We thank Dr. M. Su for communication and the anonymous referee for insightful suggestion. This work was supported in part by National Natural
Science of China under grants 10925315, 10973041, 11073057
 and 11105155. YZF is also supported by the 100 Talents program of Chinese
Academy of Sciences. QY acknowledges the support from the Key Laboratory
of Dark Matter and Space Astronomy of Chinese Academy of Sciences.\\


\end{document}